\documentstyle[12pt]{article}
\newcommand{\ce}{{{\cal E}}}
\newcommand{\cek}{{{\mathrm k}}}
\newcommand{\Uf}{{{\mathrm U}}}
\newcommand{\nf}{{{\mathrm n}}}
\newcommand{\Pf}{{{\mathbf P}}}
\newcommand{\Bf}{{{\mathbf B}}}
\newcommand{\Ef}{{{\mathbf E}}}
\newcommand{\rf}{{{\mathbf r}}}
\begin{document}

\title{ Evolution equations for eigenvalues and coefficients of polynomials and related generalized dynamics }

\author{
Robert M. Yamaleev\\
Facultad de Estudios Superiores,\\
Universidad Nacional Autonoma de Mexico\\
Cuautitl\'an Izcalli, Campo 1, C.P.54740, M\'exico.\\
Joint Institute for Nuclear Research, Dubna,Russia\\
 Email:iamaleev@servidor.unam.mx }
 \maketitle
%
% Abstract
%
\begin{abstract}
Invariant theory as a study of properties of a polynomials under translational transformations is developed. Class of polynomials with congruent
set of eigenvalues is introduced. Evolution equations for eigenvalues and coefficients remaining the polynomial within proper class of
polynomials are formulated. The connection with equations for hyper-elliptic Weierstrass and hyper-elliptic Jacobian functions is found.
Algorithm of calculation of eigenvalues of the polynomials based on these evolution equations is elaborated.
 Elements of the generalized dynamics with $n$-order characteristic polynomials is built.

\end{abstract}


\section{Introduction}

The problem of expressing of eigenvalues of the polynomials as a certain functions of the coefficients is one of the oldest mathematical
problems. The question on possibility, or impossibility,  to express the eigenvalues of the polynomials through coefficients on making use of
radicals had been exhaustively answered by E.Galois and H.Abel  that the polynomial higher than fourth order, in general, does not admit a
presentation of solutions via radicals \cite{History}. In spite of this rigorous mathematical theory mathematicians remained to believe that the
eigenvalues of the polynomials can be expressed in analytical way as certain functions of the coefficients \cite{Belard}, \cite{Lajtin}. Hermite
was first who found a very elegant expression of eigenvalues of the quintic equation by modular functions \cite{Hermite}. The theory of elliptic
functions originally was related with the problem of finding of eigenvalues of the cubic polynomial. In fact, the Weierstrass elliptic functions
at the periods are equal to eigenvalues of the cubic equation \cite{Weber}. It is clear, however, that for a search of analytical solutions of
the $n>5$-degree polynomials one needs of mathematical tools beyond the elliptic functions. In this context as a hopeful tools one may take the
theories of hyper-elliptic functions \cite{Mordell} and multi-complex algebras \cite{Lipatov}, \cite{Yamaleev1}.

The main purpose of the present paper is to construct evolution equations for eigenvalues and coefficients of polynomials. We search such kind
of evolution equations which remain the original polynomial within certain class of polynomials. Therefore first of all we define the set of
invariants and classify the polynomials with respect to the obtained set of invariants. The polynomials from the same class possess with
congruent set of eigenvalues, i.e. inside a given class the eigenvalues of one polynomial are obtained by simultaneous translations of the
eigenvalues of the other polynomial. The algorithm of calculation of the eigenvalues is based on the evolution process reducing the number of
the coefficients of initial polynomial. During of the evolution the initial polynomial will transformed into the other polynomial remaining
within the frames of a given class. The evolution is directed in such a way that the final polynomial will possess with one trivial solution.
The coefficients of the finial polynomial are solutions of the Cauchy problem for ordinary differential equations where the coefficients of the
polynomial serve as initial data. As soon as the Cauchy problem is resolved, the eigenvalues of initial polynomial are found simply by certain
set of translations of the eigenvalues of the final polynomial. It is shown, if the coefficients obey to equations for Weierstrass
hyper-elliptic functions then the eigenvalues will obey equations for hype-elliptic Jacobi functions.

The present method has been procreated in the process of construction of generalized electrodynamics of $n$-th order (see,
Refs.\cite{Yamaleev2},\cite{Yamaleev3},\cite{Yamaleev4}). In this theory an important role plays the fact that the mapping between inner- and
outer-momenta is built as a mapping between coefficients and eigenvalues of the characteristic $n$-degree polynomial. The generalized dynamics
is based on dynamic equations of motion previously developed for the polynomials. As a starting platform of the construction served the elements
of the relativistic dynamics closely related with quadratic polynomial.

Besides the Introduction the paper contains the following  sections.

In Section 2, the equations of evolution for the coefficients of $n$-degree polynomial are formulated. In Section 3, the Algorithm of finding of
eigenvalues of the corresponding polynomial is built. In Section 4, some peculiarities  of the cubic equation is explored. In Section 5, the
elements of the relativistic dynamics based on quadratic characteristic polynomial is presented. In Section 6 we give an account of a sketch of
the generalized dynamics related with $n$-degree characteristic polynomial.

\section{Evolution equations for eigenvalues and coefficients of $n$-degree polynomial }

If $F$ is a field and $q_1,...,q_n$ are algebraically independent over $F$, the polynomial
$$
p(X)=\prod_{i=1}^n(X-q_i)
$$
is referred to as {\it generic polynomial} over $F$ of degree $n$. The polynomial equation of $n$-degree over field $F$ is written in the form
$$
p(X):=X^{n}+\sum_{k=1}^{n-1}(-)^k(n-k+1)P_kX^{n-k}+(-)^{n}P^{2}=0, \eqno(2.1)
$$
where the coefficients $P_k\in F(q_1,...,q_n)$. In this paper we shall restrict our attention only to polynomials with real coefficients and
with simple roots. Moreover the last coefficient $P^2$ is essentially positive. The signs at the coefficients in Eq.(2.1) are changed from term
to term which allows in Vi\`{e}ta formulae to keep only the positive signs. The expressions at the coefficients are included purely for
convenience, and have no real bearing on the theory. The mapping from the set of eigenvalues onto the set of coefficients is given by Vi\`{e}ta
formulae
$$
(a)~~nP_1=\sum_{i=1}^{n}q_i,~~(b)~~P^{2}=\prod_{i=1}^{n}q_i,~~(c)~~P_k=\sum_{1\leq r_1<\ldots<r_k\leq n}\prod^k_{j=1}q_{r_j}, \eqno(2.2)
$$
here $P_k$ is called the {\it elementary symmetric $k$-degree polynomials} of the eigenvalues. The number of monomials inside $k$-th elementary
polynomial is equal to binomial coefficient:
$$
C^k_n=\left( \begin{array}{c}
k\\
n
\end{array} \right)=
\frac{n!}{k!(n-k)!}. \eqno(2.3)
$$
Since the roots of the generic polynomial $p(X)$ are algebraically independent, this polynomial is, in some sense, the most general polynomial
possible.

In $p(X)$ replace $X$ with $X=Y+P_1$. This transformation will eliminate the $(n-1)$-degree term, resulting in a polynomial of the form
$$
r(Y):=Y^{n}+\sum_{k=2}^{n-1}(-)^kR_kY^{n-k}+(-)^nR_0=0. \eqno(2.4)
$$
The polynomials $p(X)$ and $r(Y)$ have the same splitting field and hence the same Galois group. Let $E$ be the splitting field of $r(Y)$, let
$y_k,k=1,..,n$ be its roots in $E$ and $G=G_F(E)$ be its Galois group.

{\bf Lemma 2.1}

{\it The coefficients $R_k,~R_0$ of Eq.(2.4) are invariants with respect to simultaneous translations of the eigenvalues of Eq.(2.1)}

The {\bf Proof} of the statement it follows from formula $Y=X-P_1$ which in terms of the eigenvalues is expressed as follows
$$
y_k=q_k-\frac{1}{n}\sum_{i=1}^{n}q_i=\frac{1}{n}\sum_{i\neq k}^n (q_k-q_i). \eqno(2.5)
$$
It is seen that the eigenvalues of Eq.(2.4) are represented by differences between the eigenvalues of Eq.(2.1), hence, they are invariants with
respect to the simultaneous translations. Since the coefficients $R_0,R_k,k=2,...,n-1,$ are sum of uniform monomials of $y_k,k=1,...,n$  they
have same feature, namely, they are invariants with respect to simultaneous translations of the eigenvalues of Eq.(2.1), too.

{\bf End of Proof.}

The polynomial $r(Y)$ we denominate as {\it invariant polynomial} (with respect to translations of the roots of (2.1)).

The main task of this section is to introduce evolution equations for the coefficients of Eq.(2.1) which remain invariant the coefficients of
Eq.(2.4) $R_k$. This result is given by the following

{\bf Theorem 2.2}

{\it Let $q_{k}, k=1,...,n$ be set of eigenvalues of polynomial equation of $n$- degree (2.1). Let the differentials of all eigenvalues are
equal to each other
$$
dq_1=dq_2=\ldots=dq_k=\ldots=dq_n.\eqno(2.6)
$$
Then differentials of the coefficients satisfy the following system of equations:}
$$
2P_{n-1}dP_1=dP^2,~\eqno(2.7)
$$
$$
dP_{n-k}=(k+2)P_{n-k-1}~dP_1,~~k=1,...,n-2; \eqno(2.8)
$$

{\bf Proof.}

Notice that from (2.2a) it follows that
$$
dq_k=dP_1,~k=1,...,n.
$$
Coefficients of the polynomial (2.1) are symmetric forms of the eigenvalues where $k$-th coefficient $(n-k+1)P_k$ consists of $C^{k}_n$
monomials of $k$-degree. Thus, the derivative of this coefficient $(n-k+1)dP_k$  contains $kC^{k}_n$ monomials and, since the derivatives  all
eigenvalues are equal to each other, so, the derivative of $(n-k+1)P_k$ is proportional to $\lambda dP_1$ where the coefficient of
proportionality consists of $kC^{k}_n~$ $(k-1)$-degree symmetric monomials. On the other hand, the symmetric polynomial of $(k-1)$-th degree can
expressed only by $C^{k-1}_n$ symmetric monomials of $(k-1)$-degree. This means the expression for $\lambda$ consists of
$kC^{k}_n/C^{k-1}_n=n-k+1$ same symmetric polynomials of $(k-1)$-degree which are equal to the $(k-1)$-th coefficient  $(n-k+2)P_{k-1}$. The
result is expressed as follows
$$
(n-k+1)dP_k=(n-k+1)(n-k+2)P_{k-1}~dP_1,~~k=2,3,...,n-1.
$$
Differentiation of the last coefficient gives
$$
dP^2=2P_{n-1}dP_1, \eqno(2.9)
$$
which completes the system of differential equations for the coefficients of Eq.(2.1).

{\bf End of Proof.}

The following Lemma demonstrate an important role of the invariant polynomial in the evolution process.

{\bf Lemma 2.3}

{\it The first integrals of evolution equations (2.8) are given by coefficients of invariant polynomial (2.4).}

{\bf Proof}

In fact, the roots of Eq.(2.4), $y_k,~k=1,...,n$, according to formulae (2.5), are the first integrals of evolution equations for eigenvalues
(2.6). Equations (2.7)-(2.9) are consequences of Eqs.(2.6), hence coefficients $R_k$ as algebraic functions of the solutions of evolution
equations are first integrals of Eqs.(2.7)-(2.9).

{\bf End of Proof}

Inversely, the use formula $Y=X-P_1$ in Eq.(2.4) has to transform invariant equation into equation (2.1). Substitute $X-P_1$ instead of $Y$ and
gather together powers of $X$ and the coefficients of the obtained polynomial compare with coefficients of Eq.(2.1). We shall see that  the
coefficients $P_k,~k=1,...,n-1$ now are expressed as $k$-degree polynomials of $P_1$ with coefficients consisting of $R_k,~k=2,...,n-1$.
Especially notice, the invariant $R_0$ is defined by $n$-degree polynomial of $P_1$ with coefficients built from $R_k$. The first task is to
find an explicit form of this polynomial. The general form of those polynomial can be presented as follows
$$
P_1^{n}+\sum_{k=2}^{n-1}P_1^{n-k}f_k(R_1,...,R_k)+R_0=P^2.
$$
Now, the task is to find an explicit form of the function $f_k$. With that purpose explore firstly the case $P^2=0$. In this case one of the
solutions of Eq.(2.1) is equal to zero. Then the corresponding solution of the invariant polynomial is
$$
y_1=-P_1(P^2=0)=-P_1(0).
$$
Hence $(-P_1(0))$ will satisfy (2.4). By replacing $Y$ with $(-P_1(0))$ in Eq.(2.4) we come to the following equation for $P_1$:
$$
P_1^{n}(0)+\sum_{k=2}^{n-1}R_k P_1^{n-k}(0)+R_0=0. \eqno(2.10)
$$
Notice, here the signs at all coefficients are positive. Secondly, suppose that we have changed coefficients of Eq.(2.1) from the set
$\{~P_k,~P^2\neq 0~\}$ to the set $\{~\widetilde{P}_k,~P^2=0~\}$ obeying evolution equations (2.8). This way of evolution provides the
polynomial with $P^2=0$ with the same invariants as the original one. Hence,  when $P^2\neq 0$ the coefficients of Eq.(2.10) will not change,
but now this polynomial will be equal to $P^2$:
$$
P_1^{n}+\sum_{k=2}^{n-1}R_k P_1^{n-k}+R_0=P^2. \eqno(2.11)
$$
Notice, the situation is somewhat similar the conventional {\it Classical Invariant Theory of Polynomials} \cite{Olver}. If classical invariant
theory is a study of properties of a polynomial $p(x)$ that are unchanged under fractional linear transformations, within the framework of the
present approach we study properties of polynomials under translational transformations.

{\bf Theorem 2.4}

{\it Let coefficients of Eq.(2.1) obey evolution equations (2.7)-(2.9). Then }
$$
(k+1)P_{n-k}=\frac{1}{k!}\frac{d^k}{{dP_1}^k}P^2.\eqno(2.12)
$$

{\bf Proof}.

 Differentiate equation (2.11) by taking into account that $R_0,R_k,k=1,...,n-1$ are constants. We get
$$
dP_1~(~nP_1^{n-1}+\sum_{k=2}^{n-2}(n-k)R_k P_1^{n-k-1}+R_{n-1}~)=dP^2. \eqno(2.13)
$$
Compare this equation with equation (2.9). It is seen, the expression inside brackets at $dP_1$ in (2.13) is nothing else than the coefficient
$2P_{n-1}$ expressed as a polynomial of $P_1$ with invariant coefficients:
$$
2P_{n-1}=nP_1^{n-1}+\sum_{k=2}^{n-2}(n-k)R_k P_1^{n-k-1}+R_{n-1}. \eqno(2.14)
$$
Hence,
$$
2P_{n-1}=\frac{d}{{dP_1}}P^2.
$$
Next, differentiate (2.14) with respect to $P_1$, we obtain
$$
dP_{n-1}=3P_{n-2}~dP_1,\eqno(2.15)
$$
where $3P_{n-2}$ in fact is the next coefficient of Eq.(2.1):
$$
3P_{n-2}=~\frac{n(n-1)}{2}P_1^{n-2}+\sum_{k=2}^{n-3}\frac{(n-k)(n-k-1)}{2}R_k P_1^{n-k-2}+R_{n-2}.\eqno(2.16)
$$
Hence,
$$
3P_{n-2}=\frac{d^2}{{dP_1}^2}P^2.
$$
At the next step we shall obtain
$$
dP_{n-2}=4P_{n-3}~dP_1,\eqno(2.17)
$$
where the expression at the differential $dP_1$ we denoted by $4P_{n-3}$ because this expression indeed is $(n-3)$-th coefficient of Eq.(2.1):
$$
4P_{n-3}=~\frac{n(n-1)(n-2)}{1\cdot 2\cdot 3}P_1^{n-3}+\sum_{k=2}^{n-4}\frac{(n-k)(n-k-1)(n-k-2)}{1\cdot 2\cdot 3}R_k P_1^{n-k-3}+R_{n-3}.
\eqno(2.18)
$$
Hence,
$$
4P_{n-3}=\frac{d^3}{{dP_1}^3}P^2.
$$
From these formulae by induction one may easily establish that the general formula for $l$-th coefficient $P_{n-l}$ is
$$
(l+1)P_{n-l}= \left( \begin{array}{c}
l\\
n
\end{array} \right)
P_1^{n-l}+\sum_{k=2}^{n-l-1} \left( \begin{array}{c}
l\\
n-k
\end{array} \right)
R_k P_1^{n-k-l}+R_{n-l}=\frac{1}{l!}\frac{d^l}{{dP_1}^l}P^2. \eqno(2.18)
$$

{\bf End of Proof}.

This theorem has some interesting consequences.

{\bf Corollary 2.5}

{\it The following representation for polynomial $p(X)$ holds true}
$$
\exp(-X\frac{d}{dP_1})P^2=0. \eqno(2.19)
$$

{\bf Proof}

 The Euler operator, generator of translation, is
represented by following expansion
$$
\exp(-X\frac{d}{dP_1})= 1-X \frac{d}{d P_1}+X^2\frac{1}{2!}\frac{d^2}{d P_1^2}+...+ X^n\frac{1}{n!}\frac{d}{d P_1^n}+...~.  \eqno(2.20)
$$
By differentiating Eq.(2.11) $n$-times we get
$$
\frac{d}{d P_1^n}P^2=n!. \eqno(2.21)
$$
Hence the sum in (2.20) is completed by this term. Thus,
$$
\exp(-X\frac{d}{dP_1})P^2=p(X). \eqno(2.22)
$$
Here the variable $X$ means one of the roots of Eq.(2.1) in quality of which let us take $q_n$.

The derivative with respect to $P_1$ due to (2.6) can be expressed by the following sum
$$
\frac{d}{dP_1}=\sum_{k=1}^n\frac{\partial q_k}{\partial P_1}\frac{\partial}{\partial q_k}= \sum_{k=1}^n\frac{\partial}{\partial q_k}.
\eqno(2.23)
$$
Then,
$$
\exp(-q_n\frac{d}{dP_1})P^2=\prod_{k=1}^{n-1}\exp(-q_n\frac{\partial}{\partial q_k})\exp(-q_n\frac{\partial}{\partial q_n}).\eqno(2.24)
$$
Now, take into account Vi\`{e}ta  formula for $P^2$ given by (2.2). The Euler operator (2.24) will translate each root by $q_n$. The last
operator in (2.24) acts only upon $n$-th root resulting $q_n-q_n=0$. Hence,
$$
\exp(-X\frac{d}{dP_1})P^2=p(X)=0.
$$

{\bf End of Proof}.

\section{ Algorithm of finding the eigenvalues of $n$-degree polynomials}

 The main idea of the present algorithm is to reduce the problem of solution of
$n$-degree polynomial equation into the problem of solution of $n-1$ degree polynomial equation. The evolution from $n$-degree polynomial up to
$(n-1)$-degree polynomial is fulfilled in such a way that remains invariant coefficients $R_k$. Hence, initial and final polynomials of this
evolution will possess with congruent set of eigenvalues, so that the solutions of the former can be obtained from the solutions of the latter
simply on making use of transformations of translation.

One of the ways to reduce a degree of the polynomial is achieved
by tending $P^2$ to zero. For that purpose we must to use $P^2$ as
an evolution parameter of the evolution equations (2.8) with final
goal to find  the coefficients of Eq.(2.1) at the point $P^2=0$.
This evolution will remain invariant the coefficients of Eq.(2.4),
hence from solutions of the polynomial with $P^2=0$ we may come to
the solutions of the original equation simply by translation of
the set of solutions with new coefficients.

Denote $x=P^2$. Re-write Eq.(2.8) with respect to $x$. We get
$$
2P_{n-1}{dP_1}=dx,~~,
$$
$$
dP_{n-k}=(k+2)P_{n-k-1}dP_{1},~~k=1,...,n-3; \eqno(3.1)
$$
$$
\frac{dP_2}{dx}=nP_1dP_{1}.
$$

This a well-known {\it Cauchy problem} with initial data $P_k(x=P^2),~~k=1,2,3,...,n-1$. The variable $x$ run from $x=P^2$ till $x=0$. These
equations usually are resolved by using the celebrated {\it Cauchy-Lipschitz} method of calculation \cite{Cauchy}. This procedure is fulfilled
by dividing the interval $(x_0,0)$ into $N$ parts:
$$
\Delta x_0=x_1-x_0,~\Delta x_i=x_{i+1}-x_i,~\Delta x_{n-1}=x-x_{n-1},\eqno(3.2)
$$
where $x_i<x_{i+1},~x_N=0.$

 In this way the continuous evolution process is transformed into discrete process consisting of $N$ steps. In the last step of
we come to $n$-degree polynomial free of the last coefficient:
$$
{\stackrel{(1)}{X^n}}+ \sum_{k=1}^{n-1}(-)^k(n-k+1){\stackrel{(1)}{P_k}} {\stackrel{(1)}{X^k}}=0.\eqno(3.4)
$$
 One of the solutions is trivial, excluding this solution we come to the polynomial of $(n-1)$-degree:
$$
{\stackrel{(1)}{X^{n-1}}}+ \sum_{k=1}^{n-1}(-)^k(n-k+1){\stackrel{(1)}{P_k}} {\stackrel{(1)}{X^{k}}}+2{\stackrel{(1)}{P_{n-1}}}=0. \eqno(3.5)
$$
Let us mention that Eq.(3.4) possesses with same invariants as original one, i.e. Eq.(2.1). If one will use the numerical methods of solution
then the relationships for invariants are satisfied within given accuracy of calculations. Suppose that the solutions of $(n-1)$-degree equation
(3.5) ${\stackrel{(1)}{q}}_k,~k=1,...,n-1$ are known. Complete this set of solutions by ${\stackrel{(1)}{q}}_n=0.$ Then the solutions of
original equation one may find simply by the following translations
$$
q_k={\stackrel{(1)}{q_k}}+P_1-{\stackrel{(1)}{P_1}},~k=1,...,n.\eqno(3.6)
$$
If solutions of the $(n-1)$-degree equation still are unknown, then one may continue to apply this algorithm again in order to reduce the
problem of solution of $(n-1)$-degree equation to the problem of solution $(n-2)$-degree polynomial equation. This process can be continued up
till quadratic or linear equation. At each step of iteration one will find an information on coefficient ${\stackrel{(r)}{P}}_1$. At the $r$-th
iteration one deals with $(n-r)$-degree equation in the form
$$
{\stackrel{(r)}{X^{n-r}}}+ \sum_{k=1}^{n-r-1}(-)^k(n-k+1){\stackrel{(r)}{P_k}} {\stackrel{(r)}{X^k}}+(r+1){\stackrel{(r)}{P_{n-r}}}=0.
\eqno(3.7)
$$
At this stage the norm of the coefficient ${\stackrel{(r)}{P}}_{n-r}$ will fulfil the role of evolution parameter of the next evolution process.
As soon as the  solutions of the lowest polynomial is found, the inverse process of iterations will consists only translations of the known set
of eigenvalues with known parameter of translation. Let the last step of iteration be linear equation from which we find only one solution,
$$
{\stackrel{(n-1)}{q_1}}=n{\stackrel{(n-1)}{P_1}}.
$$
Then the solutions of the original equation (2.1) are found as a result of the following set of translations
$$
q_1=n{\stackrel{(n-1)}{P_1}}+n\sum_{r=2}^n\frac{1}{r}(~{\stackrel{(n-r)}{P_1}}-{\stackrel{(n-r+1)}{P_1}}~),\eqno(3.8a)
$$
$$
q_s=\sum_{r=s}^n\frac{1}{r}(~{\stackrel{(n-r)}{P_1}}-{\stackrel{(n-r+1)}{P_1}},~s=2,3,...,n;\eqno(3.8b)
$$
where ${\stackrel{(0)}{P_1}}=P_1$.

\section{ Relativistic Lorentz-force equations and related quadratic polynomial }

Inter-relation between the evolution of the eigenvalues an the coefficients of the polynomials with dynamic equations for physical systems we
can observe already at the level of quadratic polynomial. In aim of this section is to demonstrate as the quadratic polynomial is related with
relativistic Lorentz-force equations. In the sequel, we shall use this example as a starting platform to pass to case of higher degree
polynomials and related generalized dynamics. Let us start with generic quadratic polynomial
$$
X^2-2P_0~X+P^2=0, \eqno(4.1)
$$
with real coefficients $P_0^2\geq P^2$, and real eigenvalues $p^2_1,p^2_2$. This polynomial closely related with the relativistic dynamics
\cite{Yamaleev2}, \cite{Gal}. In order to demonstrate this fact let us start from Lorentz-force equations for a charged particle inside external
electromagnetic fields $\Ef,~\Bf$ written with respect to proper time $\tau$ {\footnote{
 Hereafter for the sake of simplicity we omit all parameters like charge, mass, light-velocity and other parameters regularizing physical
 dimensions taking them equal to unit.}}
$$
\frac{d\Pf}{d\tau}=(\Ef~P_0+[\Pf\times \Bf]),~ \frac{dP_0}{d\tau}=(\Ef \cdot \Pf), \eqno(4.2)
$$
$$
\frac{d\rf}{d\tau}=\Pf,~~\frac{dt}{d\tau}={P_0}. \eqno(4.3)
$$
Consider projection of equations (4.2) on direction of the motion defined by unit vector $\vec n=\Pf /P$:
$$
\frac{d P}{d\tau}=(\Ef~\cdot \nf)~P_0,~ \frac{dP_0}{d\tau}=(\Ef \cdot \nf)~P, \eqno(4.4)
$$
Then one deals only with the lengths of the momenta $P_0,~P$. Simplify Eqs.(4.4) by introducing new evolution parameter:
$$
\frac{dP}{ds}=P_0,~~\frac{dP_0}{ds}=P, \eqno(4.5)
$$
where
$$
\frac{ds}{d\tau}=(\Ef\cdot \nf). \eqno(4.6)
$$
The first constant of motion is easily found
$$
P_0^2-P^2=M^2. \eqno(4.7)
$$
Here $M^2$ is a constant of motion. Conventionally, this constant is interpreted as a {\it square of inertial mass} \cite{Adan}.
 Inside stationary potential field, when $\Ef=-{\nabla}{V(r)}$, dynamic equations imply the other integral of motion, the energy,
$$
\ce_0=P_0+V(r). \eqno(4.8)
$$
At the rest state where $P=0$ one obtains $P_0(P=0)=M$. The relativistic mechanics deals with two kinds of the energy, namely,
$$
p^2_1=P_0-M,~~p^2_2=P_0+M.\eqno(4.9)
$$
From formulae (4.7) and (4.9) it follows
$$
P=p_1p_2,~P_0=\frac{1}{2}(p_2^{2}+p_1^{2}),~ M=\frac{1}{2}(p_2^{2}-p_1^{2}). \eqno(4.10)
$$
Notice, the first two formulae of (4.10) mean Vi\`{e}ta formulae for quadratic polynomial equation (4.1). Substitution $X=Y+P_0$ in (4.1) leads
to {\it invariant equation}
$$
Y^2=P_0^2-P^2=M^2,\eqno(4.11)
$$
the invariant coefficient of which is equal to the invariant of physical motion.  Quadratic equation for $P_0$ is given by
$$
P_0^2-M^2=P^2.\eqno(4.12)
$$
Differentiating this equation we derive evolution equations for the coefficients, which in the case of quadratic polynomial, of course, is a
simple task:
$$
2P_0~P=\frac{d}{ds}P^2,~~~\frac{d}{ds}P_0=P.\eqno(4.13)
$$
Solutions of these evolution equations are given by hyperbolic cosine-sine functions
$$
P=M~sinh(s),~~P_0=M~cosh(s). \eqno(4.14)
$$
Then the eigenvalues of Eq.(5.1) are expressed by hyperbolic cosine-sine functions of one-half argument
$$
p_1^2=\sqrt{2M}~sinh(\frac{s}{2}),~~p_2^2=\sqrt{2M}~cosh(\frac{s}{2}). \eqno(4.15)
$$

Now, consider so-called the {\it effective potential representation}. With this purpose let us come back to the equations written with respect
to proper-time $\tau$. Then from the second equation of (4.13) it follows
$$
\ce_0=P_0+V(r).
$$
Further, replace $P_0$ by $\ce_0-V(r)$ in the first of equation of (4.13), this gives
$$
\frac{d}{ds}\Pf=-\nabla V(r)~(\ce_0-V(r)=-\nabla W(r,\ce_0),\eqno(4.16)
$$
where the effective potential is defined by
$$
W(r\ce_0)=\ce_0V(r)-\frac{1}{2}V^2(r).\eqno(4.17)
$$
Here, the relativistic Lorentz-force equation written in the form of Newtonian equation with the {\it effective potential} $W$. From this
equation it follows the Newtonian form of the energy
$$
\ce =\frac{1}{2}P^2+W(r,\ce_0)=\frac{1}{2}(\ce_0^2-M^2).\eqno(4.18)
$$

\section{Cubic polynomial equation and related dynamics}

In the previous section we have demonstrated as the relativistic dynamics is related with the evolution of quadratic polynomial. This example
provides us with appropriate tool in order to construct generalized scheme based on the evolution of polynomials of higher order. In this
section we explore the case of cubic polynomial
$$ p(X)=X^{3}-3P_1~X^{2}+2P_2~X-P^{2}=0,\eqno(5.1)
$$
relations between coefficients and eigenvalues are given by
$$
3P_1=q_1+q_2+q_3,~ 2P_2=q_1q_2+q_2q_3+q_3q_1,~P^{2}=q_1q_2q_3.\eqno(5.2)
$$
We shall restrict ourselves only with the case when coefficients are represented by real numbers and let us assume that $p(X)$ is irreducible.
By replacing $X$ with $X=Y+P_1$ we come to invariant polynomial
$$
Y^{3}+R_2~Y-R_{0}=0,\eqno(5.3)
$$
where
$$
~(a)~~R_2=2P_2-3P_1^{2},~~~(b)~~P_1^{3}+R_2~P_1+R_0=P^{2}.\eqno(5.4)
$$
The eigenvalues and the coefficients of this polynomial are invariants with respect to simultaneous translations of the eigenvalues of Eq.
(5.1). Obviously this statement is a consequence of the formula $ Y=X-P_1$ from which it follows
$$
3y_1=e_2-e_3,~3y_2=e_3-e_1,~3y=e_1-e_2,~~\eqno(5.5)
$$
where
$$
e_1=q_3-q_2,e_2=q_1-q_3,e_3=q_2-q_1.
$$
The evolution equations remaining constants coefficients $R_0,R_2$ are obtained directly from Eqs.(5.4). Differentiating these equations we get
$$
dP_1~(3P^2_1+R_2)=2P_2~dP_1=dP^2,\eqno(5.6)
$$
$$
dP_2=3dP_1.
$$
Evolution equations for the eigenvalues, obviously, have to has the following form
$$
\frac{d}{ds}q_k=A,~~k=1,2,3,
$$
where $A$ is some function same for all eigenvalues. In order to construct some dynamics in the quality of $A$ we must take $A=P$. Then
$$
\frac{d}{ds}q_k=\frac{d}{ds}P_1=P.\eqno(5.7)
$$

The cubic polynomial is an object of special interest, because this polynomial is closely related with the classical elliptic functions. The
solutions of the evolution equations for the eigenvalues and the coefficients of the cubic polynomial can be represented via elliptic Jacobi and
Weierstrass functions, correspondingly \cite{Akhiezer}. On making use (5.7) from (5.4b) we come to the following differential equation
$$
P_1^{3}+R_1~P_1+R_0= (\frac{dP_1}{ds})^2 .\eqno(5.8)
$$
Write this equation in the following designations
$$
(\frac{d\wp}{dz})^2=4\wp(z)^3-g_2\wp(z)-g_3,\eqno(5.9)
$$
where $z=4s$, $g_2=-4R_1,~~g_3=-4R_0$ and $\wp(2s)=P_1(s)$. The
integral formula for $\wp(z)$ is given by
$$
z=\int^{\infty}_{\wp}({4x^3-g_2 x-g_3})^{-1/2}dx.\eqno(5.10)
$$
The functions $\wp(z)$ and $\wp'(z)$ are Weierstrass elliptic functions with periods $2\omega_1,~~2\omega_2$. Define
$\omega_3=-\omega_1-\omega_2$. Then the values
$$
\wp(\omega_1),~\wp(\omega_2),~\wp(\omega_3)$$ are the roots of
cubic equation
$$
4x^3-g_2x-g_3=0.\eqno(5.11)
$$

Introduce new variables $p_k,~k=1,2,3$ where $p^2_k=q_k$. Then $P=p_1p_2p_3$. For $p_k$ evolution equations are derived from (5.7):
$$
\frac{dp_1}{ds}=p_2p_3,~~\frac{dp_2}{ds}=p_1p_3,~~\frac{dp_3}{ds}=p_2p_1.\eqno(5.12)
$$
 Solutions of these equations  presented by quotients of Jacobi elliptic functions. Let $sn(u),cn(u),dn(u)$ be a set of Jacobi elliptic
functions. Define the following quotients of these functions (in Glaisher notations \cite{Akhiezer})
$$
ns=-\frac{1}{sn},~~cs=-\frac{cn}{sn},~~ds=-\frac{dn}{sn},
$$
which obey the following differential equations
$$
\frac{d}{du}ns(u)=cs(u)ds(u), \frac{d}{du}cs(u)=ns(u)ds(u),\frac{d}{du}ds(u)=cs(u)ns(u),\eqno(5.13)
$$
with
$$
ns^2-cs^2=1,~~ds^2-cs^2=1-\cek.
$$
From Eqs.(4.12) it follows that the values
$$
e_1=q_3-q_2,~e_2=q_1-q_3,
$$
are constants of motion. Hence solutions of Eqs.(4.12) are presented via Jacobi elliptic functions as follows
$$
q_3={e_1}ns^2(u,\cek ),~q_2={e_1}cs^2(u,\cek),~q_1={e_1}ds^2(u,\cek),~~\cek=1-\frac{e_2}{e_1}.\eqno(5.14)
$$

The final result  let us represent by the following\\

{\bf Statement}:

{\it If squared roots of the eigenvalues of the cubic equation obey equations for Jacobi elliptic functions, then the evolution of the
coefficients are governed by equations for Weierstrass elliptic functions.}

For cubic equation there exist, also, another possibility to express its eigenvalues, namely, via trigonometric functions. To the formulae given
below we come virtue of formulae (3.25) derived in Ref.\cite{Lipatov}.

The eigenvalues of the reduced polynomial are given by formulae:
$$
y_1=-\frac{2}{3}\sqrt{3R_1}\cos\theta,~y_2=\frac{1}{3}\sqrt{3R_1}(\cos\theta+\sqrt{3}\sin\theta),~
y_3=\frac{1}{3}\sqrt{3R_1}(\cos\theta-\sqrt{3}\sin\theta). \eqno(5.15)
$$
It is easy to verify that
$$
y_1+y_2+y_3=0,~~y_1y_2+y_2y_3+y_3y_1=-R_1.
$$
Equation for the last coefficient, $R_0$, leads to trigonometric equation for $\cos(\theta)$:
$$
-R_0=y_1y_2y_3
$$
$$
=\cos\theta(\cos^2\theta-3\sin^2\theta)=4\cos^3\theta-3\cos\theta=\frac{27}{2}\frac{R_0}{(\sqrt{3R_1})^{3}}.
$$
By using the trigonometric formula
$$
\cos(3\theta)=4\cos^3\theta-3\cos\theta
$$
we come to simple trigonometric equation:
$$
\cos(3\theta)=\frac{3\sqrt{3}}{2}\frac{R_0}{\sqrt{R_1^3}}. \eqno(5.16)
$$
Since we restrict ourselves only with real solutions the following inequality has to be true
$$
(\frac{R_0}{2})^2< (\frac{R_1}{3})^3. \eqno(5.17)
$$

The algorithm elaborated in the previous section in the case of cubic equation (5.1) is simplified as follows. With respect to $x=P^2$ as an
evolution parameter evolution equations (5.6)-(5.7) have a form
$$
2P_2{dP_1}={2P_2}dx,~~ 2P_2{dP_2}={3P_1}dx, \eqno(5.18)
$$
with the initial data $P_1(x=P^{2})=P_1,~P_2(x=P^{2})=P_2$. At the final stage when $x=P^2=0$ one finds
$P_1(P^2=0)={\stackrel{(1)}{P_1}},~P_2(P^2=0)={\stackrel{(1)}{P_2}}$ satisfying relationships
$$
R_2=2{\stackrel{(1)}{P_2}}-3{\stackrel{(1)}{P_1^{2}}},~~{\stackrel{(1)}{P_1^{3}}}+R_1~{\stackrel{(1)}{P_1}}+R_0=0.
$$
Thus the new equation possesses with same invariants as the original one,
$$
X^3(0)-3{\stackrel{(1)}{P_1}}X^2(0)+2{\stackrel{(1)}{P_2}}X(0)=0. \eqno(5.19)
$$
Equations (5.1) and (5.19) possess with congruent eigenvalues, however Eq.(5.19) has one trivial root. Therefore the problem is reduced to the
solution of quadratic equation
$$
X^2(0)-3{\stackrel{(1)}{P_1}}X^2(0)+2{\stackrel{(1)}{P_2}}X(0)=0. \eqno(5.20)
$$
From solutions of eq.(5.20) to the solutions of Eq.(5.1) one comes simply by the set of translations:
$$
q_1=P_1-{\stackrel{(1)}{P_1}},~~q_2=q_2(0)+P_1-{\stackrel{(1)}{P_1}},~~q_3=q_3(0)+P_1-{\stackrel{(1)}{P_1}}.
$$

The dynamics related with the cubic polynomial, evidently, is characterized with two first constants of motion $R_2,~R_0$ which do not depend of
potential field. In quality of evolution equations we have to use Eqs.(5.6)-(5.7) written with respect to parameter of evolution defined in
(4.6):
$$
P_2~=\frac{d}{ds}P,~~~\frac{d}{ds}P_2=3dP_1,~~\frac{d}{ds}P_1=P.\eqno(5.21)
$$
Evolution equations for the eigenvalues, correspondingly, are given by
$$
\frac{d}{ds}p^2_k=P,~~k=1,2,3.\eqno(5.22)
$$
From these equations it follows that the constants of motion are given by formulae
$$
M_1=p^2_2-p^2_3,~M_2=p^2_3-p^2_1,~M_3=p^2_1-p^2_2.\eqno(5.23)
$$
In order to include into this scheme the potential field $V(r)$ we have to use equation (4.6):
$$
\frac{ds}{d\tau}=(\Uf\cdot \nf),~~\Uf=\nabla V(r),~~P\nf=\Pf.
\eqno(5.24)
$$
Furthermore, the set of evolution equations has to be completed with the interrelation between momentum and  velocity with respect to time-like
parameter:
$$
\Pf=\frac{d\rf}{d\tau}.\eqno(5.25)
$$
In these designations evolution equations (5.21) are transformed
into the following dynamic equations
$$
\frac{d}{d\tau}\Pf=-\nabla
V(r)~P_2,~~\frac{d}{d\tau}P_2=-3(\Pf\cdot\nabla)V(r)
P_1,~~\frac{d}{d\tau}P_1=-(\Pf\cdot\nabla) V(r).\eqno(5.26)
$$
In the case of stationary potential, beside the first constant $R_2,R_0$, the equations imply another constant of motion, the energy
$$
\ce_1=P_1+V.\eqno(5.27)
$$
On making use of expression for $P_1$ from (5.27) in the second
equation of Eqs.(6.6), we get
$$
\ce_2=P_2+3\ce_1V-\frac{3}{2}V^2=\frac{1}{2}(R_2+3\ce_1^2).\eqno(5.28)
$$
This is second expression for the energy. Next, express from this
formula $P_2$ and substitute into first equation of (5.26). This
leads to Newtonian equation
$$
\frac{d}{d\tau}\Pf=-\nabla W(r,\ce_1),\eqno(5.29)
$$
with effective potential
$$
2W(r,\ce_1)=\ce_2V(r)-3\ce_1V^2(r)+V^3(r).\eqno(5.30)
$$
From Eq.(5.29) one may find formula for the total energy
$$
\ce=\frac{1}{2}P^2+W=\frac{1}{2}(R_0+R_2\ce_1+\ce_1^3).\eqno(5.31)
$$

\section{ Generalized dynamics related with evolution of $n$-degree polynomial}

Now we have accumulated enough experience in order to be able to build the general form of the dynamics related with $n$-degree polynomials. Let
us start with evolution equations (2.8) written for $n$-degree polynomial (2.1). Dynamic equations with respect to some time-like evolution
parameter $s$ describing a motion inside stationary potential field $V(r)$ are formulated in the following form
$$
\frac{d\Pf}{d\tau}=\Uf~P_{n-1},~~ \eqno(6.1a)
$$
$$
\frac{dP_{k}}{d\tau}=(\Uf\cdot\Pf)P_{k-1}({n-k+2}),~~ k=2,\ldots,
n-1, \eqno(6.1b)
$$
$$
\frac{dP_{1}}{d\tau}=(\Uf\cdot\Pf). \eqno(6.1c)
$$
The set of evolution equations has to be completed with the
interrelation between momentum and  velocity given by Eq.(5.25).
This system of dynamic equations remain invariant the coefficients
$R_0,R_k,k=1,...,n-1$. The motion of a physical system obeying to
these equations possesses with the set of {\it inner} and {\it
outer} momenta. Evolution of the set of outer-momenta $\{
P^2,P_k,k=1,...,n-1~\}$ are given by Eqs.(6.1), whereas the
evolution of the inner-momenta $\{ p_k,k=1,...,n\}$ is described
by
$$
\frac{dp^2_k}{d\tau}=(\Uf\cdot\Pf),~k=1,...,n.\eqno(6.2)
$$
The first integrals of this system are given by
$$
M_{ik}=p^2_i-p^2_k,~~i\neq k.\eqno(6.3).
$$

From Eq.(6.2c) we find the first constant of integration, the
energy $\ce_1=P_1+V$, or $P_1=\ce_1-V.$ By substituting $P_1$ into
the next, $k=2$-th equation of (6.2b) we find the other constant
of integration (the second expression for energy):
$$\ce_2=P_2+nE_1~V-\frac{n}{2}V^2.$$
By continuing this process, namely, by substituting $P_2$ from the last expression into the next equation with $k=3$ we shall find the third
constant of integration:
$$\ce_3=P_3-(n-1)E_2~V+\frac{n(n-1)}{2}E_1~V^2-\frac{n(n-1)}{2\cdot 3}V^3.$$
Continue this process up till $(n-1)$-th stage. Then, in the
$(n-1)$-th stage we shall obtain the expression for $P_{n-1}$. By
introducing this expression into Eq.(6.2a), we come to Newtonian
equation
$$
\frac{d}{d\tau}\Pf=-\nabla W(r,\ce_1), \eqno(6.4)
$$
where the effective potential is given by the following series
$$
W(r,\ce_1)=\frac{1}{2} \sum^n_{k=1} (k+1)\ce_{n-k}
V^k(r)(-1)^{k+1},~\mbox{ with}~ \ce_0=\frac{1}{n+1}.  \eqno(6.5)
$$
From Eq.(6.4) one may find formula for the total energy
$$
\ce=\frac{1}{2}P^2+W=\frac{1}{2}(R_0+\sum_{k=2}^{n-1}R_k\ce_1^{k-1}+\ce_1^n).\eqno(6.6)
$$

{\bf Concluding remarks}

This is a prodigious fact that the evolutions equations elaborated for polynomials serve as dynamic equations for the generalized dynamics of
high-energy particles. Notice, the Algorithm of calculation of eigenvalues of the polynomials elaborated in this paper is distinct of the
numerical methods of calculations which principally are based on iteration process with initial sampling, or tentative, data of roots, so that
effectiveness of these algorithms depends of the initial data. The iteration process based on the present Algorithm uses in the quality of
initial data the coefficients of the original polynomial. The method elaborated here can be considered also as a theory of functional connection
between coefficients and eigenvalues of the polynomial expressed as one-valued Weierstrass and Jacobi hyper-elliptic functions. Evidently, the
present method without any principal difficulties can be continued to the case of polynomials defined over the field of complex numbers.

We have restrict our attention only on dynamic equations given in one-dimensional coordinate space.  On the theory of the generalized dynamics
in $4D$ space with physical units one may consult in Refs.\cite{Yamaleev3} and \cite{Yamaleev4} and the references therein.


\begin{thebibliography}{99}

\bibitem{Akhiezer}
Akhiezer N.I., Elements of theory of elliptic functions, Moscow, "Nauka", 1970.
\bibitem{Belard}
G.Belardinelli, Fonctions hypergeometriques de plusieurs variables et resolution analitique des equations algebriques generales. Paris:
Gauthier-Villars, 1960.
\bibitem{History} F.Cajoris, History of Mathematics, New York, 1919, pp.349-350.\\
I.Richards, An application of Galois theory to elementary arithmetic, {\it Advances in Mathematics} {\bf 13} (1974) 268-273.
\bibitem{Cauchy} Harold T.Davis, Introduction to nonlinear differential and integral euations. Dover Publications, INC.,N.Y., 1968.
(ISBN 0-486-60971-5)
\bibitem{Gal} P.Fjelstad, S.Gal,
Two dimensional geometries, topologies, trigonometries and physics generated by complex-type numbers, {\it Advances in Applied Clifford
algebras}, {\bf 11(1)}, (2001), 81.
\bibitem{Hermite}
Ch.Hermite, Sur la resolution de l'eqation du cinquieme degree C.R.Acad.Sci.Paris. - 1858. V.46.\\
Ch.Hermite, Sur l'eqation du cinquieme degree C.R.Acad.Sci.Paris. - 1865. V.46; 1866.-V.62.
\bibitem{Lajtin}
L.K.Lakhtin, Algebraic equations resolvables in hypergeometric functions (in russian) M.,1893.\\
Differential resolvents of algebraic equations of higher degree. M.,1896.
\bibitem{Lipatov} L.N.Lipatov, M.Raush de Traubenberg, G.G.Volkov, On ternary complex analysis and its applications. arXiv: 0711.0809v1 [math-ph ]
6 Nov. 2007.
\bibitem{Mordell} L.J.Mordell, On the rational solutions of the indeterminate equations of the 3-rd and 4-th degrees. {\it Proc.Camb.Phill.Soc.}
{\bf 21} (1922) 179-192.
\bibitem{Olver} P.J.Olver, Classical Invariant Theory, LondonMathematical Society Student Texts 44, Cambridge Univeristy Press, 1999.
\bibitem{Adan} A.R.Rodriguez-Dominguez, Lorentz.force equations as Heisenberg equations for 4D-quantum system. {\it Revista Mexicana de Fisica}
{\bf 53} (4) (2007) 270-280.
\bibitem{Weber}
H.Weber, Lehrbuch der Algebra.Bd.2. Gruppen. Lineare Gruppen. Anwendungen der Gruppen Theorie. Algebraische Zahlen.- Braunschweig, 1899.\\
H.Weber, Lehrbuch der Algebra.Bd.3. Elliptische Functionen und algebraische Zahlen.- Braunschweig, 1908.
\bibitem{Yamaleev1} R.M.Yamaleev, Multicomplex algebras on polynomials and
generalized Hamilton dynamics, {\it J.Math.Anal.Appl.}, {\bf 322}, (2006), 815-824.\\
R.M.Yamaleev, Complex algebras on N-order polynomials and generalizations of trigonometry, oscillator model and Hamilton dynamics.  {\it
Advances in Applied Clifford algebras}, {\bf 15(1)}, (2005), 123.
\bibitem{Yamaleev2} R.M.Yamaleev, "Relativistic
Equations of Motion within Nambu's Formalism of Dynamics", {\it Ann.Phys.} {\bf 285 (2) } (2000) 141-160.
\bibitem{Yamaleev3} R.M.Yamaleev, "Generalized Newtonian Equations of Motion", {\it Ann.Phys.} {\bf 277 (1)} (1999) 1-18;\\
R.M.Yamaleev, "Elliptic and Hyperelliptic Deformed Mechanics in $n$-Dimesional Phase Space", {\it JINR
Communications},P2-94-109,Dubna, (1995);\\
R.M.Yamaleev, "Generalized Lorentz-Force Equations", {\it Ann.Phys.} {\bf 292 (1)} (2001) 157-178.
\bibitem{Yamaleev4}
R.M.Yamaleev, Extended Relativistic Dynamics of Charged Spinning Particle in Quaternionic Formulation, {\it Advances in Applied Clifford
Algebras}, {\bf 13(2)}, (2003), 183-218.\\
R.M.Yamaleev, Ternary Electrodynamics, {\it Far East J.Dynamic Systems}, {\bf 9(2)} (2007) 307-324.



\end{thebibliography}
\end{document}